\documentclass[a4paper,11pt]{article}
\usepackage{graphicx}%
\usepackage{float}
\usepackage{geometry}
\usepackage{cite}
\makeatletter
\def\@cite#1#2{\textsuperscript{#1\if@tempswa , #2\fi}}
\makeatother

\begin{document}
 \title{Group delay of electromagnetic pulses through multilayer dielectric mirrors beyond special relativity}
\author{Jiang-Tao Liu$^{1,2*}$\footnote{$^{1}$Department of Physics, Nanchang University, Nanchang
330031, China. $^{2}$Institute for Advanced Study, Nanchang University, Nanchang 330031, China. $^{3}$Department of Physics, Semiconductor Photonics Research Center, Xiamen University, Xiamen 361005,
China. $^{4}$Key Laboratory of Materials Physics, Institute of Solid State Physics, Chinese Academy of Sciences
Hefei 230031, China. $^{*}$jtliu@semi.ac.cn. $^{\dag}$fhsu@issp.ac.cn.}, Wu Xin$^{1}$, Nian-Hua Liu$^{1,2}$, Jun Li$^{3}$ and Fu-Hai Su$^{4\dag}$ }

\maketitle

\textbf{A number of unexpected phenomena historically broke through corresponding the traditional physics framework, thus motivating scientists to develop more advanced and accurate theories. A well-known example is quantum tunnelling in $\alpha$-decay. The quantum tunnelling rate can be described accurately by the quantum theory but cannot be explained by classical mechanics. In quantum tunnelling, the length of time during which a particle  tunnels through a barrier remains undetermined\cite{1CEU,2MLA,3HTE,4HEH,5LR,6WHG}.  Based on classical quantum and classical electrodynamics theories, the  group delay during quantum tunnelling is independent of barrier thickness. However, in more accurate theories, such as the general theory of relativity\cite{7EA}, is this condition still valid? In this Letter, we investigate the group delay of optical pulses through multilayer dielectric mirrors (MDM) combined with gravitational wave (GW)\cite{7EA,8WJM,9LIGO,10AP}. We find that the delay increases linearly with MDM length for the transmitted wave packet, whereas the Hartman effect disappears. Our study provides insight into the nature of both  quantum tunnelling and GW.}

The length of time during which quantum particles tunnel through a barrier has attracted considerable
attention for both fundamental and technological reasons since the 1930s\cite{1CEU,2MLA,3HTE,4HEH,5LR,6WHG}. Hartman calculated the tunneling of a wavepacket through a rectangular potential barrier\cite{3HTE} and found that group delay becomes constant as barrier length increases. This phenomenon, known as the Hartman effect, implies that for sufficiently large barriers, the effective group velocity of a particle may be superluminal. Although a number of experiments have reported observations of electromagnetic waves propagating with "superluminal tunneling velocities," the definition of tunneling time and its exact physical meaning based on experimental results remain under heated debate\cite{4HEH,5LR,6WHG}. A large number of tunneling time definitions have been proposed, including group delay or phase time\cite{3HTE}, dwell time\cite{6WHG}, Larmor times\cite{11BAI,12RVF}, and B\"{u}ttiker-Landauer time\cite{13BM}.  Winful recently proposed that the group delay in tunneling represents a lifetime of stored energy escaping through both sides of the barrier and does not represent a transit time\cite{6WHG, 14WHG}. Thus, the issues of superluminality, causality, or the speed of information transfer do not even arise.

Beyond Schr\"{o}dinger's nonrelativistic quantum mechanics, the group delay for Dirac particles traveling through a potential well was also studied by using Dirac's fully relativistic quantum theory\cite{15KP}. The behavior of Dirac particles is found to be the same  as that in nonrelativistic quantum mechanics. Liu et al. recently studied the  effect of the electromagnetic fields of the group delay of electrons and found that the group delay of the transmitted wave packet increases linearly with barrier length for the transmitted wave packet\cite{16LJT}. This peculiar tunneling effect is attributed to current leakage in a time-dependent barrier generated via the electromagnetic fields\cite{13BM,16LJT}. If the quantum fluctuation or the zero-point field is considered, all potential barriers are combined with electromagnetic fields. Thus, in the framework of quantum field theory, the Hartman effect of electron tunneling may disappear because of the existence of electromagnetic vacuum fields.

However, in photon tunnelling, a number of theories based on the special relativistic covariant Maxwell equations  have proven that the group delay of photons becomes constant as the length of optical structures increases\cite{4HEH,5LR,6WHG, 17SAM,18SSC,19CJJ,20LCF}. However, whether the Hartman effect still exists when a more accurate theory (e.g. general relativity theory) remains unclear. One of the unique predictions in general relativity theory is the existence of gravitational wave (GW)\cite{7EA,8WJM,9LIGO,10AP}. If  optical structures such as  a multilayer dielectric mirror (MDM) (i.e. a one-dimensional photonic crystal) is irradiated by GW, the center frequency and the width of the photonic bandgap of the MDM will vary with the GW. Similar to the electron tunnelling in a time-dependent barrier, variations of the photonic bandgap will result in an additional leak photon current. Such current may propagate at the speed of light. Thus, the Hartman effect may also be absent in photon tunnelling within the framework of general relativity theory.

In this paper, we investigate the effect of GW on the group delay of photon passing through MDM. Our simulation
shows that with a thick MDM, group delay increases linearly with increasing barrier width. The group velocity is approximately $2.95\times10^{8}$ m/s, slightly less than the speed of light in in vacuum. Superluminality or causality  no longer occurs.  We also find that the group delay of tunneling photons is sensitive to GW. Our study may facilitate further understanding of both quantum tunnelling and GW as well as  provide a different method for the detection of GW.  In particularly, the MDM comprises alternating dielectric layers and vacuum layers [see Fig. 1(a)]. All layers are nonmagnetic ($\mu=1$), and the thicknesses of the dielectric layers (vacuum layers) satisfy $D_{1}=\lambda_{0}/4\sqrt{\varepsilon_{1}}$ ($D_{2}=\lambda_{0}/4+\zeta_{2}\lambda_{0}/2$), where $\varepsilon_{1}$ is the permittivity of dielectric layers, $\lambda_{0}$ is the center frequency of the input electromagnetic pulse, and $\zeta_{2}$ is a positive integer. The group delay of tunneling photons is generally more sensitive to GW at large $\zeta_{2}$. We set $\zeta_{2}=5$ in this paper, unless otherwise specified. The electromagnetic pulse is incident along the normal of the surface of MDM, and the propagation direction of the plane-polarized GW is parallel to the surface of MDM.

When the polarization GW occurs, the layer spacing of dielectric layers will vary with the GW.  Similar to the case of two masses separated by a distance D along the Z direction that are coupled by a lossy spring, the equation for the differential motion of the masses in the GW becomes\cite{7EA,21WJ,22WR}
\begin{equation}
\frac{d^{2}z_{R}}{dt^{2}}+\frac{\omega _{m0}}{Q}\frac{dz_{R}}{dt}+\omega
_{m0}^{2}z_{R}=\frac{1}{2}\frac{d^{2}h_{22}}{dt^{2}}D,  \label{eqn1}
\end{equation}
where $z_{R}$ is the proper relative displacement of the two masses, $h_{22}$ is the the perturbation matrix (tensor) element resulting from the GW; and $\omega _{m0}$ and $Q$ are the natural frequency and the q-factor of the lossy spring, respectively. If the dielectric layers in MDM are free and under zero-velocity initial conditions, the equation of differential motion can be described  by\cite{7EA,21WJ,22WR}
\begin{equation}
z_{R}/{D}=h_{22}/2=A_{GW}\cos(\omega_{GW}t)/2,  \label{eqn2}
\end{equation}
where $A_{GW}$ is the GW amplitude.  Thus, in the propagation of electromagnetic waves, the permittivity distribution will also change with time. To study such a time-dependent photon scattering process, we employ the finite-difference time-domain (FDTD) method to solve the time-dependent Maxwell equations numerically\cite{23KY,24AT}. FDTD is a time-domain technique that can  provide animated displays of electromagnetic field movement through various  models, e.g., the photonic crystals, radar, etc.  In the FDTD method, the one-dimensional Maxwell equations are replaced by a finite set of finite differential equations\cite{23KY,24AT}
\begin{equation}
\left\{
\begin{array}{c}
E_{x}^{n+1}(k)=E_{x}^{n}(k)-\frac{\Delta t}{\varepsilon \Delta z\Upsilon_{GW}}%
\left[ H_{y}^{n+1/2}(k+1/2)-H_{y}^{n+1/2}(k-1/2)\right],  \\
H_{y}^{n+1/2}(k+1/2)=H_{y}^{n-1/2}(k+1/2)-\frac{\Delta t}{\mu \Delta z\Upsilon
_{GW}}\left[ E_{x}^{n+1}(k+1)-E_{x}^{n}(k)\right],
\end{array}%
\right.\label{eqn3}
\end{equation}
where  $E_{x}$ ($H_{y}$) is the  electric field (magnetic field) of the electromagnetic wave, $(i, k) = (i\triangle x, k\triangle t)$ denote a grid point of the space and time; and for any function of space and time $F(i\triangle x, k\triangle t) =F^{k}(i)$, $\Upsilon_{GW}$ depends on the GW wave-induced displacement. According to Eq. \ref{eqn2}, in the vacuum layers $\Upsilon_{GW}=1+A_{GW}\cos(\omega_{GW}t)/2$. The thickness changes of the dielectric layers are minimal because the natural frequency of dielectric layer is nonresonant with the frequency of GW, and the thickness the dielectric layers is considerably less than that of the vacuum layers with large $\mathcal{L}_{2}$. Thus, we set $\Upsilon_{GW}=1$ in dielectric layers. At the input boundary, a Gaussian electromagnetic wave packet is injected. The wave function at the input boundary is set as $E_{x}=H_{y}=\frac{1}{\sqrt{2}}\exp \left[ -4\pi\left( t-\tau _{0}\right)
^{2}/\tau _{0}^{2}\right] e^{i\omega _{0}t}$. To reduce  distortion and  numerical errors, a relatively long pulse is used: $\tau _{0}=200 T_{0}$, where $T_{0}$ is the period of the electromagnetic wave.

By numerically solving Eq. (\ref{eqn3}) directly in the time domain, the propagation of a wave packet through a barrier can be demonstrated in real time. For computational stability, the space increment $\Delta x$ and
the time increment $\Delta t$ need to satisfy the relation $\Delta x>c \Delta t$. Furthermore, the space increment $\Delta x$ must be significantly smaller than the wavelength of electromagnetic wave $\Delta x<\lambda_{0}/12$, and the time increment $\Delta t$ must be considerably smaller than the period of the GW. To ensure high precision, the space increment $\Delta x=\lambda_{0}/1.5 \times 10^{3}$ and the time increment $\Delta t=2 \times 10^{-4} T_{0}$ are used. When the space and time increments are increased or reduced 10 times, the error is less than 3\%.

\begin{figure}[t]
\centering
\includegraphics[width=0.98\columnwidth,clip]{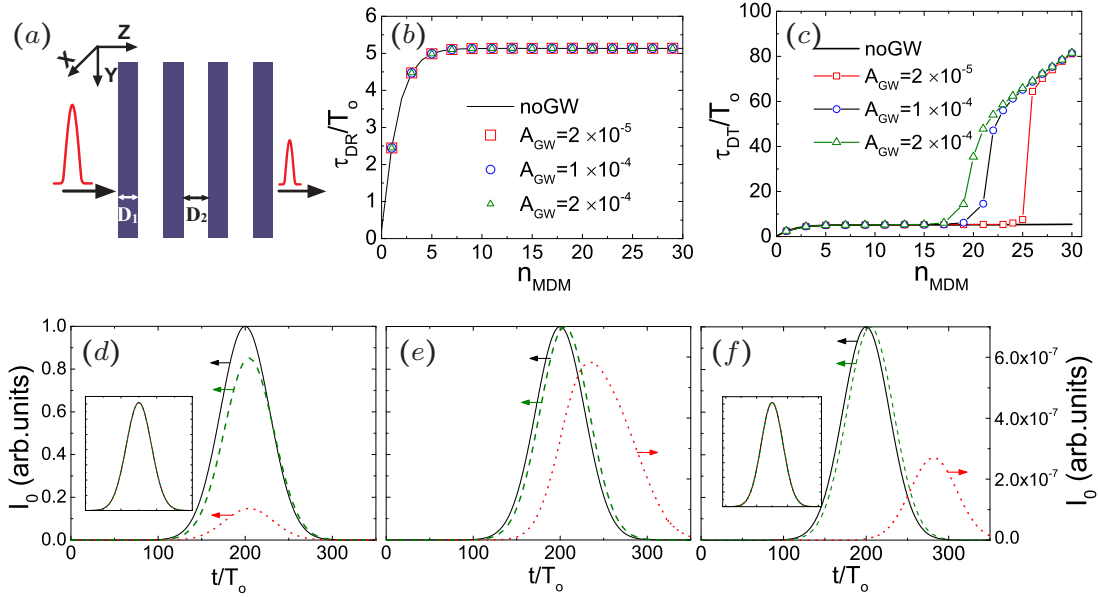}
\caption{(a) Schematic diagram of the tunneling process in an MDM structure. (b) Group delay for the reflected wave packet and (c) that
for the transmitted wave packet as a function of the number of MDM periods with different GW amplitudes.
Incident (black lines), tunneled (red lines) and reflected (green lines) pulses with GW amplitude
$A_{GW}=1\times10^{-4}$ for the following number of MDM periods: (d)$n_{MDM}=3$, (e)$n_{MDM}=19$, and (f) $n_{MDM}=30$.
The inset shows the normalized tunneled pulse overlaid with the incident pulse.}%
\label{fig1}%
\end{figure}

Numerical results of the group delay (i.e., the delays of the peaks of the reflected and transmitted pulses) are shown in Figs. 1(b)-(f). Fig. 1(b) shows the group delay for the reflected wave packet as a function of the number of MDM periods $n_{MDM}$. Similar to the traditional quantum tunneling, the group delay is saturated by increasing $n_{MDM}$, and the saturated group delay is identical to the dwell time. Meanwhile, the group delay for the reflected wave packet is unaffected by the extrinsic GW.

However, the influence of the GW on the group delay of a transmitted wave packet differs from the given case. As shown in Fig. 1(c), for the GW amplitude $A_{GW}=1\times 10^{-4}$,  the group delay for the transmitted wave packet increases linearly when $n_{MDM}>24$.  This result can be explained by the variations of the photonic bandgap  attributed to the GW. Similar to the electron tunnelling in a time-dependent barrier\cite{13BM,16LJT}, the variations of the photonic bandgap will result in an additional leak photon current. Given that the amplitude of the additional leakage current attributed GW is quite small, for the case of a small number of MDM periods, e.g., $n_{MDM}<15$,  the tunneling current is significantly larger than the additional leakage current and is unaffected by the extrinsic GW [see Fig. 1(c)]. However, for the case of large number of MDM periods, e.g., $n_{MDM}>24$, the tunneling current is significantly weaker than the additional leakage current. The additional leakage current determines the group delay, and the group delay increases with the amplitude of GW. Under a weaker GW, the additional leakage current will determine the group delay for the transmitted wave packet with a relatively larger  $n_{MDM}$.   Specifically, for $A_{GW}=1\times 10^{-4}$ ($A_{GW}=2\times 10^{-5}$), the group delay increases linearly when $n_{MDM}>24$ ($n_{MDM}>26$). Thus, if $n_{MDM}$ is sufficiently large, even under a relatively weak GW (e.g., the GW background radiation\cite{25BGJ}), no ¡®Hartman effect¡¯ occurs. On the other hand, for a non-strictly periodic GW emitted by various sources(e.g., the chaos compact binary system\cite{26ZSY}), the time-dependent variations attributed to the GW will also result in an additional leak photon current, thus modifying the group delay.   From Fig. 1(c), we can also find that the group velocity of the additional leakage current is independent of the amplitude of GW. The group velocity is approximately $2.95\times 10^{8}$ m/s, which is the same as the speed of light in vacuum. No superluminal appears.

The transmitted wave packet includes two components: the tunneling current and the additional leakage current induced by the time-dependent modulation, which may distort the transmitted wave packet. However, the relative weight of the contribution of the tunneling current and that of the additional leakage current vary with the number of MDM periods $n_{MDM}$. For a small number of MDM periods, e.g., $n_{MDM}=3$, the amplitude of the tunneling current is significantly larger than that of the additional leakage current. Similar to in traditional quantum tunneling, the distortion is minimal [see Fig. 1(d)]. As the number of MDM periods increases, the tunneling rates decrease rapidly. Meanwhile, the number of MDM periods only has a slight effect on  the additional leakage current. For a large number of MDM periods, e.g, $n_{MDM}=19$,  the amplitude of the additional leakage current and that of the tunneling current are comparable. Considering that the group delay of the additional leakage current is larger than that of the tunneling current, a serious distortion of transmitted wave packet occurs [see Fig. 3(e)]. However, if the number of MDM periods is sufficient, e.g., $n_{MDM}=30$,  the tunneling rate is very small, and the additional leakage current is the main contributor to the transmitted wave packet. No distortion occurs at this scale [see Fig. 3(f)].

However, we have to determine whether the group delay of the undistorted transmitted wave packet in a thick MDM with GW equates to tunneling time. In traditional quantum tunneling, the consensus is that the group delay does not equate to a tunneling time  for the following reasons: the group delay is equal to the dwell time, which denotes a lifetime of stored energy  escaping through the barrier, not a pulse traveling  through the barrier\cite{6WHG,14WHG}; to reduce the distortion, the length of the wave packet is often significantly larger than the gap length of the barrier\cite{6WHG,27AT}, and the peak of the input pulse does not even propagate into the barrier. However, the group delay of photons with GW is different. The group delay with GW cannot be explained by the dwell time because such delay is considerably larger than the dwell time. The full width at half maximum (FWHM) of the injected wave packet is approximately $66\lambda_{0}$ (or $66T_{0}$), which is smaller than the total optical path length of MDM at $n_{MDM}=30$.  Meanwhile, the group delay for $n_{MDM}=30$ is approximately $80T_{0}$, which is larger than the FWHM of the injected wave packet. The peaks of the injected wave packet and that of the transmitted wave packet are distinguishable. Thus, the group delay of the transmitted wave packet in a thick MDM with GW may be regarded as the tunneling time. However, although the group delay equates to the tunneling time, the tunneling time is increased with the number of MDM periods. Thus, the superluminality in the tunneling process no longer occurs.

\begin{figure}[t]
\centering
\includegraphics[width=0.6\columnwidth,clip]{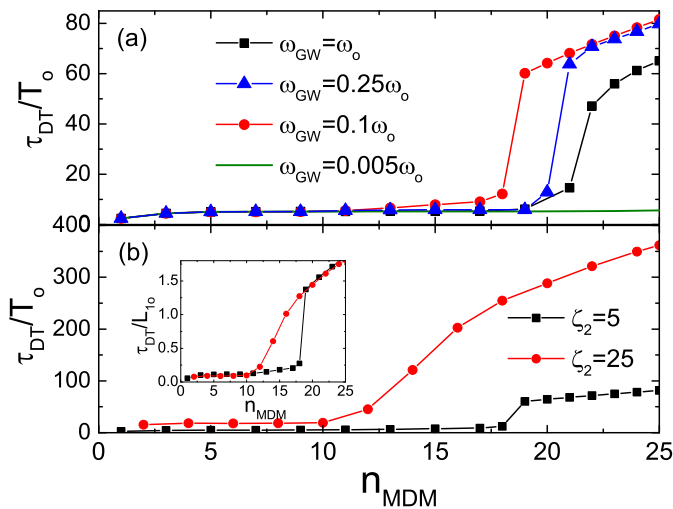}
\caption{(a) Group delay of the transmitted wave packet as a function of the number of MDM periods for different
GW frequencies with $A_{GW}=1\times 10^{-4}$. (b) Group delay of the transmitted wave
packet as a function  of the number of MDM periods for different vacuum layer thicknesses
with $A_{GW}=1\times 10^{-4}$, $\omega_{GW}=0.1\omega_{0}$, and $\tau _{0}=2000 T_{0}$. The inset shows the relative group delay as
a function  of the number of MDM periods.}%
\label{fig2}%
\end{figure}

Notably, the group delay in the tunneling process also shows good sensitivity to the GW. For instance, for $n_{MDM}=25$ and $A_{GW}=1\times 10^{-4}$, the group delay without  (with) GW is approximately $5.3T_{0}$ ($65T_{0}$).  The group delay is increased by approximately 12 times. In the  GW detection by using a Michelson interferometer, the interferometer measures the intensity rather than the time delay of the interference light. We can still make a comparison with Michelson interferometer. For a Michelson interferometer with an arm length of $L_{o}=75\lambda_{0}$ (same as the total optical path length of MDM with $n_{MDM}=25$), only when the GW amplitude $A_{GW}$ is approximately $1.6\times 10^{-3}$  (i.e., the GW amplitude satisfies $4\pi A_{GW}L_{o}/\lambda_{0}\approx \pi/2$), the intensity of the interference light with and without GW can vary by approximately 12 times.

The sensitivity of the group delay to the GW depends on the frequency of the GW [see Fig. 2(a)]. For high-frequency GW, e.g., $\omega_{GW}=\omega_{0}$, the variations of the photonic bandgap occur rapidly, the stored electromagnetic energy in MDM changes slightly, and the additional leak photon current is relatively weak. The group delay increases remarkably only when $n_{MDM}>22$. However, for a relatively low-frequency GW, the additional leak photon current is enhanced.  The group delay increases remarkably when $n_{MDM}>21$ ($n_{MDM}>19$) for $\omega_{GW}=0.25\omega_{0}$ ($\omega_{GW}=0.1\omega_{0}$). For  an extremely low-frequency GW, e.g., $\omega_{GW}=0.01\omega_{0}$, the period of the GW  becomes larger than the FWHM of electromagnetic waves, and the effect of the GW on the group delay becomes small.  For $n_{MDM}<25$, the group delay does not change significantly.

The sensitivity  also depends on the thickness of the vacuum layers. For MDM with thicker vacuum layers (i.e., large $\zeta_{2}$), the GW-induced variation of layer spacing is enhanced, and a larger additional leak photon current can be achieved. As shown in Fig. 2(b), group delay increases remarkably when $n_{MDM}>12$ for $\zeta_{2}=25$. However, for $\zeta_{2}=5$,  group delay increases remarkably only when $n_{MDM}>19$.  On the other hand, the relative group delay $\tau_{DT}/L_{1o}$ of the additional leakage current is independent of $\zeta_{2}$, where $L_{1o}$ is the optical path length of each MDM period [see the inset of Fig. 2(b)], which indicates that the pulses propagate with the same group velocity for different $\zeta_{2}$.

Finally, we discuss the experimental realization of our theoretical predication. Although a strong GW is used in the numerical calculation, our results show that the group delay of the tunneling photons is sensitive to GW and that sensitivity can be increased for a relatively low-frequency GW or for thick vacuum layers. Owing to ultrafast laser technology, an extremely high time resolution can be achieved. Thus, the detection of the  effect of the GW on group delay may be feasible. Notably, a pulsar emits a beam of electromagnetic radiation with ultra-high accuracy and stability. Thus, when the pulsar electromagnetic radiation  tunnels  through a strong GW radiation source such as the black hole binary, the variation of the group delay should be detectable through astronomical observation.

In conclusion, we have calculated the  group delay of optical pulses through MDM combined with GW. We found that the group delay increases linearly with MDM length for the transmitted wave packet. The Hartman effect disappears. This peculiar tunneling effect is attributed to the additional current leakage attributed to the GW-induced variations of the photonic bandgap. We also show that the group delay of the tunneling photons is sensitive to GW. For a relatively low-frequency GW or thick vacuum layers,  the sensitivity can be enhanced remarkably. Our study  provides insight into the nature both of the quantum tunnelling and GW as well as a novel process by which to detect the GW.

This work was supported by the NSFC Grant Nos. 10904059, 11004199, 11104232, 11173012, and 11264030, the NSF from the Jiangxi Province Nos. 20122BAB212003.

\end{document}